\documentclass[aps,twocolumn,pre,showkeys,showpacs,eqsecnum]{revtex4-1}
\usepackage{graphpap}
\usepackage[dvips]{graphicx}
\usepackage[dvips]{graphics}
\usepackage{color}
\usepackage{amsmath, amssymb, amsfonts, amsthm}
\usepackage[normalem]{ulem} 
\usepackage {ulem} 

\begin{document}

\title{Back action of graphene charge detectors on graphene and\\ carbon nanotube quantum dots}
\author{C. Volk$^{1,2}$, S. Engels$^{1,2}$, C. Neumann$^{1,2}$, and C. Stampfer$^{1,2}$}
\affiliation{
$^1$JARA-FIT and II. Institute of Physics, RWTH Aachen University, 52074 Aachen, Germany\\
$^2$Peter Gr\"unberg Institute (PGI-9), Forschungszentrum J\"ulich, 52425 J\"ulich, Germany}

\date{ \today}



\begin{abstract}
We report on devices based on graphene charge detectors (CDs) capacitively coupled to graphene and carbon nanotube quantum dots (QDs). We focus on back action effects of the CD on the probed QD. A strong influence of the bias voltage applied to the CD on the current through the QD is observed. Depending on the charge state of the QD the current through the QD can either strongly increase or completely reverse as a response to the applied voltage on the CD. To describe the observed behavior we employ two simple models based on single electron transport in QDs with asymmetrically broadened energy distributions of the source and the drain leads. The models successfully explain the back action effects. The extracted distribution broadening shows a linear dependency on the bias voltage applied to the CD. We discuss possible mechanisms mediating the energy transfer between the CD and QD and give an explanation for the origin of the observed asymmetry.
\end{abstract}

\maketitle

\section{Introduction}
Charge sensors have become an important tool to precisely detect localized charges in nanoelectronic systems. A common technique to realize a charge detection scheme in two-dimensional electron gases (2DEGs) is placing a quantum point contact (QPC) close to the structure carrying charges. This technique has been successfully used to realize e.g. coherent spin and charge manipulation~\cite{Petta2005,Hayashi2003} or time resolved charge detection~\cite{Gustavsson2006,Fujisawa2006}. Moreover, QPC based charge detection is a well-established technique to read out spin qubits in III/V heterostructure 2DEGs~\cite{Bluhm2010,Nowack2011,Shulman2012,Elzermann2003}. It has further been shown that the detection fidelity can be enhanced by substituting the QPC for a single electron transistor (SET) which features a higher charge to no-charge signal ratio~\cite{Goetz2008}.\\
The described detection schemes are even applicable to the relatively novel field of carbon based nanostructures, i.e. carbon nanotube~\cite{Shorubalko2008,Choi2012} and graphene quantum dots (QDs)~\cite{Guettinger2008,Volk2013}. Due to their weak spin-orbit \cite{Huertas-Hernando2006,Min2006} and hyperfine interaction \cite{Trauzettel2007}, these materials promise long-living spin states making them potentially interesting for quantum computational applications~\cite{Trauzettel2007}. To detect the charge state of carbon nanostructures, narrow graphene ribbons have been successfully integrated in carbon nanostructures and its functionality has been proven by a number of experiments~\cite{Guettinger2008,Volk2013,Engels2013}.\\
The electronic transport through graphene nanoribbons can be described by a stochastic Coulomb blockade, i.e. a series of QDs induced by edge and surface disorder~\cite{Sols2007,Stampfer2011}. Here, the slopes of the Coulomb resonances can be used to detect charging events in nearby quantum dots. This approach has been employed to perform charge sensing on individual graphene QDs~\cite{Guettinger2008,Volk2013,Fringes2011,Wang2010,Mueller2012}, including time resolved detection~\cite{Guettinger2011a}.\\
While graphene QDs together with integrated charge detectors can be fabricated out of a single graphene sheet in a single lithography step, charge sensing in carbon nanotubes is more difficult to realize. In the past, the most common approach was to fabricate close-by metallic SETs~\cite{Goetz2008,Biercuk2006} or to couple a CNT based SET via a metal gate~\cite{Zhou2012,Churchill2009}. More recently, an all-carbon hybrid device has been demonstrated where charge states in a CNT have been detected by the current through a nearby graphene nanoribbon-based charge sensor~\cite{Engels2013}.\\
Beside its ability to detect charge states, an equally important property of a charge detector is its influence on the quantum dot itself. This so-called back action is crucial and needs to be understood if the detector is used in a system where a quantum dot has to be precisly controlled (e.g. in quantum computational schemes).\\
Here, we report on the fabrication and characterization of graphene nanoribbon-based charge sensors integrated in graphene and carbon nanotube quantum dot devices. Effects of back action and counter flow are studied which give a clear indication of an energy transfer from the charge detectors to the quantum dots.

\section{Device fabrication}
\begin{figure}[]
\centering
\includegraphics [width=0.9\linewidth] {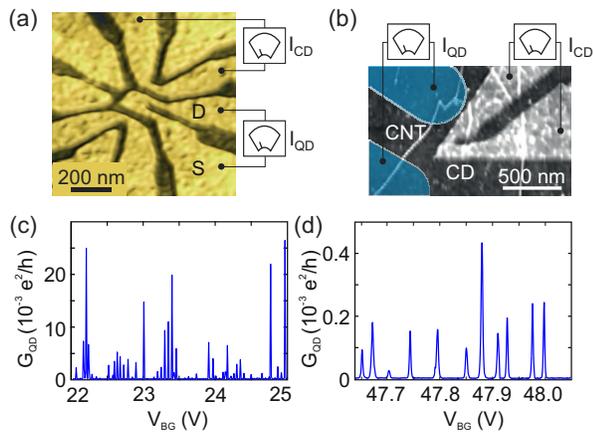}
\caption[fig01]{(color online) (a) Scanning force microscopy (SFM) image of a device consisting of a graphene quantum dot (QD) connected by graphene source (S) and drain (d) leads. Two charge detectors (CD) are realized by graphene nanoconstrictions and positioned in the vicinity to the QD. (b) SFM image of a carbon nanotube (CNT) with a nearby graphene nanoconstriction acting as a CD. The blue areas indicate the position of the contacts to the CNT. The measurement circuits are indicated by the schematic drawings in (a) and (b). (c) Conductance of the graphene QD as a function of the back gate voltage $V_{BG}$ in the Coulomb blockade regime showing distinct Coulomb resonances. (d) Measurement of Coulomb resonances of the CNT QD. }
\label{fig01}
\end{figure}

The fabrication of the device consisting of a graphene quantum dot with a nearby graphene charge detector is based on mechanical exfoliation of natural graphite. The flakes are deposited on highly doped Si$^{++}$ wafers with a 290~nm thick SiO$_2$ layer. Single-layer graphene flakes are identified by Raman spectroscopy. Individual graphene flakes are subsequently nanostructured by electron beam lithography (EBL) and reactive ion etching (RIE) using an Ar/O$_2$ plasma. The devices are contacted by an additional EBL step followed by metal evaporation. Fig.~\ref{fig01}(a) shows a scanning force microscope (SFM) image of the investigated device. The graphene QD is located in the center of the image and has a diameter of around 130 nm. The QD is connected to source (S) and drain (D) leads via narrow graphene constrictions acting as tunable tunneling barriers. The charge detectors (CDs) are realized by two nanoribbons measuring a width of about 70~nm and positioned 50~nm from the QD. Apart from detecting charges the CDs can also be used as lateral gates if a voltage is applied with respect to the QD. Additionally, the underlying Si$^{++}$ substrate can be operated as a global back gate tuning the overall Fermi energy of the device.\\
Fig.~\ref{fig01}(b) shows a SFM image of our second device scheme which consists of a CNT QD with a nearby graphene CD. To fabricate these CNT/graphene hybrid devices the CNTs are grown on Si$^{++}$ /SiO$_2$ substrates by chemical vapor deposition (CVD) using a ferritin-based iron catalyst method~\cite{Durrer2008}. This technique allows controlling the density of CNTs to be approx. 1 to 2 CNTs per $\mu m^2$. The single-walled CNTs have a typical diameter of around 1.5 to 2~nm and measure up to several micrometers in length.
After the CNT growth, graphene is deposited on the same substrates by mechanical exfoliation. Graphene flakes in close vicinity to a CNT are selected and patterned into nanoribbons by EBL and RIE. Subsequently, they are contacted as described above. In the particular device investigated in this study (see Fig.~\ref{fig01}(b)) the contacts are 350~nm apart. Furthermore, the graphene constriction acting as the CD is located at a distance of 150~nm from the CNT and is 100~nm wide.

\section{Measurements and discussion}

\begin{figure}[]
\centering
\includegraphics [width=0.9\linewidth] {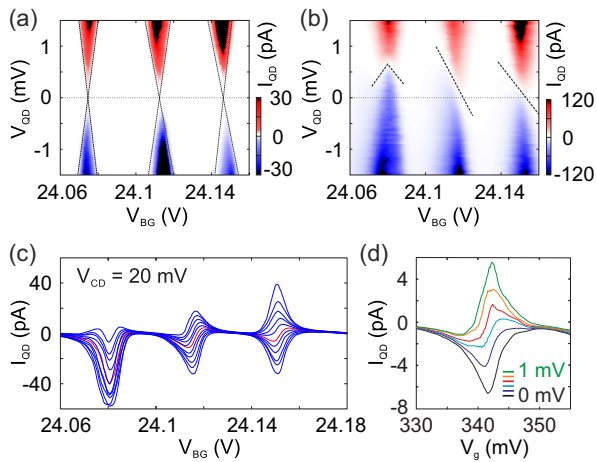}
\caption[fig02]{(color online)
(a) and (b) Current through the graphene QD $I_{QD}$ as function of $V_{QD}$ and $V_{BG}$ measured at $V_{CD}=0~$V and $V_{CD}=20~$mV, respectively. (c) Line cuts through plot (b) at bias voltages of $V_{QD} = 0, \pm0.1, \pm0.2, \pm0.3, \pm0.4$ and $\pm0.5$~mV. The zero at $V_{QD}$~=~0~mV is highlighted in red.
(d) Coulomb peak of the CNT QD in dependence of a voltage applied to the CD with respect to the QD $V_{G}$ measured at $V_{BG}=46.2$~V, $V_{CD}~=~100$~mV and applied bias voltages $V_{QD}$~=~0~mV to 1~mV as indicated in the panel.}
\label{fig02}
\end{figure}

Low temperature transport measurements are performed in a dilution refrigerator (at a temperature $T < 100$~mK) and a pumped $^4$He cryostat ($T \approx 1.5$~K). Home-built low-noise DC amplifiers (amplification factor of $10^8$, bandwidth $<1$~kHz) and low-frequency lock-in techniques are used to measure the current through the devices. Prior to the study of back action effects, suitable voltages are applied to the global back-gate and to the local lateral gates to tune the device into the Coulomb blockade regime. Please see Ref.~\cite{Volk2013,Engels2013} for further details. Figs.~\ref{fig01}(c) and 1(d) show the conductance of the graphene (c) and CNT (d) QD in dependence of the voltage applied to the back gate ($V_{BG}$). Distinct Coulomb peaks are visible which occur at every charging event of the corresponding QD. Fig.~\ref{fig02}(a) shows the current through the graphene QD as a function of the applied bias voltage ($V_{QD}$). The observed diamond shaped areas of suppressed current constitute another distinct feature of the Coulomb blockade. Please note that the functionality of the nearby charge detectors of the particular devices shown in this study have been investigated in great detail and the results have been published in Refs.~\cite{Neumann2013,Volk2013,Engels2013}.\\
To study back action effects, we repeat the measurements of Coulomb peaks (cf. Figs.~\ref{fig01}(c) and 1(d)) and bias spectroscopy (cf. Fig.~\ref{fig02}(a)) for a finite bias voltage applied to the CDs. Fig.~\ref{fig02}(b) shows a bias spectroscopy measurement at $V_{CD}$~=~$20$~mV in the same range as shown in Fig.~\ref{fig02}(a).
We observe an overall increase of the transmission through the QD for $V_{CD}$~=~20~mV which is in agreement with earlier observations of back action effects in comparable devices~\cite{Neumann2013}. Moreover, for some peaks the transition between positive and negative current moves away from the the line of $V_{QD}$~=~0~mV (see dashed lines in Fig.~\ref{fig02}(b)). For a detailed analysis Fig.~\ref{fig02}(c) shows corresponding line cuts at $V_{QD}$~=~$0, \pm0.1, \pm0.2, \pm0.3, \pm0.4$ and $\pm0.5$~mV. Here, the red trace illustrates the data set for $V_{QD}$~=~$0$~mV. Evidently, the interaction with the CD (or back action of the CD) can result in current $I_{QD}$ flowing against the biased direction which is most prominent for the left most peak. This effect is further enhanced at higher $V_{CD}$. A similar behavior can be observed for the CNT based device as shown in Fig.~\ref{fig02}(d). Here, the current through the CNT QD $I_{QD}$ is shown for $V_{CD}$~=~100~mV. For $V_{QD}$~=~0~mV $I_{QD}$ is negative and continuously develops into a positive peak as $V_{QD}$ is increased to 1~mV.\\

\begin{figure}[]
\centering
\includegraphics [width=0.9\linewidth] {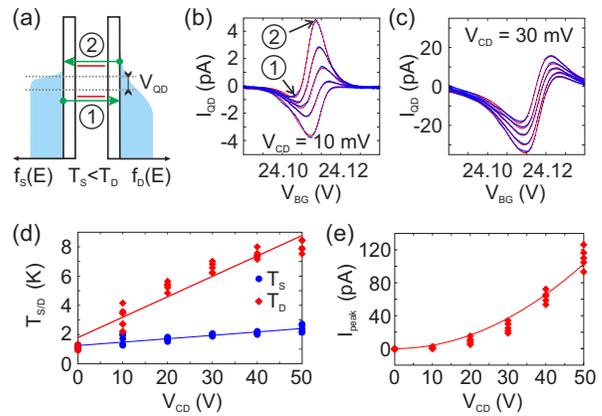}
\caption[fig03]{(color online)
(a) Illustration of the model assuming an asymmetric broadening of the Fermi distributions $f_S(E)$ and $f_D(E)$ in the leads and only considering a single ground state contributing to electronic transport. The two red lines indicate possible positions of QD levels and the corresponding direction of net current is indicated by the green arrows. (b) and (c) Comparison of experimental data (blue) and fits (red) according to the model assuming single level transport. Data for bias voltages $V_{QD}$ ranging from -0.4 mV to +0.4 mV are shown in each panel. The bias applied to the charge detector $V_{CD}$ measures 10 and 30~mV, respectively.
(d) Extracted broadening parameters $T_S$ (blue) and $T_D$ (red) as function of $V_{CD}$. Multiple data points per $V_{CD}$ value correspond to different measurements in a bias regime $|V_{QD}| \leq 0.4$~mV. The solid lines represent linear fits to the data.
(e) Absolute peak current $I_{peak}$ extracted from the negative shoulder of each resonance (see arrow labeled by (1) in (b). The solid line represents a quadratic fit.}
\label{fig03}
\end{figure}

In order to discuss the observed behavior we employ a simple model of transport through quantum dots derived from rate equations considering only one ground state in the QD:
\begin{flalign*}\label{eq:Counterflow_Thermal}
I &= \Gamma \left[f_S(E)-f_D(E)\right] &\\
& = \frac{\Gamma}{1+e^{\frac{\alpha(V_0-V_G)+V_{QD}/2}{k_B T_S}}}-\frac{\Gamma}{1+e^{\frac{\alpha(V_0-V_G)-V_{QD}/2}{k_B T_D}}}.&
\end{flalign*}
Here, $\Gamma$ is the overall tunneling rate of the tunnel barriers and $f_S(E)$, $f_D(E)$ are the Fermi distributions of the source and drain lead, respectively. Moreover, $V_0$ is the center of the Coulomb peak, $\alpha$ the gate lever arm, $V_G$ the applied gate voltage and $V_{QD}$ the bias voltage. The parameters $T_S$ and $T_D$ describe the broadening of the Fermi functions $f_S(E)$, $f_D(E)$ related to the two leads. A corresponding schematic of the model is shown in Fig.~\ref{fig03}(a). Here, the Fermi distributions $f_S(E)$ and $f_D(E)$ are assumed to exhibit a different amount of broadening with $T_S$~$<$~$T_D$ and are offset by a positive bias $V_{QD}$. This directly results in two scenarios where the gate voltage can either tune the QD state to a chemical potential $E$ where $f_S(E)$-$f_D(E)$~$>$~0 (scenario 1) or $f_S(E)$-$f_D(E)$~$<$~0 (scenario 2). Scenario 1 thus results in an enhanced flow of electrons in bias direction, while in scenario 2 the electrons flow against the bias direction.\\
Figs.~\ref{fig03}(b) and 3(c) show fits (red traces) of the described model to the experimental data (blue data points) at $V_{CD}$~=~10~mV (b) and $V_{CD}$~=~30~mV (c). Evidently, the model successfully reproduces the entire shape of the measured Coulomb resonances for different QD biases. In particular, the two scenarios 1 and 2, where the current is either enhanced (1) or flows against the direction of the applied bias (2) can be nicely explained (see e.g. arrows and labels in Fig.~\ref{fig03}(b)). To obtain more quantitative results, we fit the model to data sets measured for different $V_{CD}$ and extract the parameters $T_S$ and $T_D$. Figure~\ref{fig03}(d) summarizes the results as a function of $V_{CD}$.  We observe roughly a linear dependency of both effective temperatures $T_S$ and $T_D$ on $V_{CD}$. However, while $T_D$ only increases from $\approx$~1.0~K at $V_{CD}$~=~0~mV to $\approx$~2.5~K at $V_{CD}$~=~50~mV, $T_S$ shows a strong increase from $\approx$~1.0~K to $\approx$~8~K within the same range. This finding clearly shows the increasing asymmetry of the Fermi distribution broadening of both leads.\\
In addition, we analyzed the maximum current flowing through the QD. Fig.~\ref{fig03}(e) shows the absolute peak current extracted from the negative shoulder of each Coulomb resonance centered around $V_G \approx 24.11~$V as function of $V_{CD}$. A quadratic dependency of the peak current $I_{peak}$ on the detector bias is observed. This finding suggests that the current induced in the QD is approx. proportional to the electrical power dissipated in the CD.\\

However, the model described above has a number of shortcomings. In particular it does not provide the expected electron temperature for zero CD bias, nor does it adequatly describe all observed transport phenomena. For example, Coulomb peaks exhibiting a total inversion of the current such as the one shown in Fig.~\ref{fig02}(c) at $V_{BG}$~$\approx$~$24.08$~V or Fig.~\ref{fig02}(d) cannot be explained within the limits of the single level transport model.
Thus, all these make it difficult to rely on and fully justify the use of this oversimplified model.

\begin{figure}[]
\centering
\includegraphics [width=0.9\linewidth] {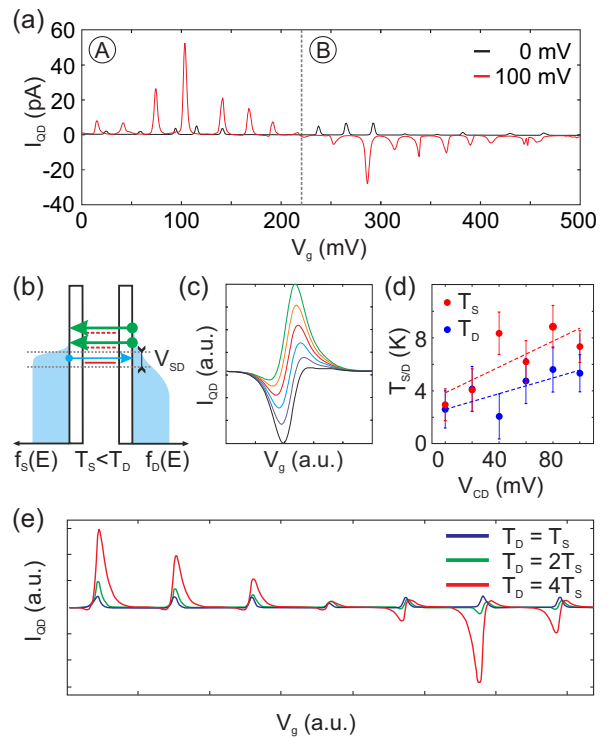}
\caption[fig04]{(color online)
(a) Current through the CNT $I_{QD}$ in dependence of $V_{G}$ at $V_{QD}=0.3$~mV, $V_{BG}=46.2$~V, $V_{CD}=0$~mV (black curve) and 100~mV (red curve). The behavior of the Coulomb peaks ranges from strong enhancement (regime A) to strong reversal (regime B) of the current.
(b) Illustration of a model taking into account three states with different tunnel coupling (see different width of green and blue arrows) contributing to transport and an asymmetric broadening with $T_S$~$<$~$T_D$.
(c) Calculations according to the model illustrated in (b) for bias voltages ranging from 0 to 1~mV.
(d) Extracted broadening parameters $T_S$ (blue) and $T_D$ (red) of the CNT QD as function of $V_{CD}$. The solid lines represent linear fits to the data.
(e) Calculated current through a QD system with broadening parameters $T_S$ and $T_D$ considering energy dependent tunneling rates which fluctuate by over 4 orders of magnitudes.
}
\label{fig04}
\end{figure}

Additional data, which cannot be described by our model is illustrated in
Fig.~\ref{fig04}(a). Here, we show a sequence of Coulomb peaks of the CNT QD recorded in a gate voltage ($V_G$) range of 500~mV for $V_{CD}$~=~0~mV (black trace) and $V_{CD}$~=~100~mV (red trace). Biasing the charge detector with $V_{CD}=100$~mV leads to an enhancement of the current in regime A ($V_G$~$<$~220~mV) by approximately one order of magnitude. In regime B ($V_G$~$>$~220~mV) a reversal of the current occurs. Additionally, the Coulomb peak heights in both regimes are modulated by an envelope function with a period of roughly 300~mV.\\
For the discussion of the observed behavior we employ a slightly more advanced model which considers three levels [i.e. the ground state (GS) and two excited states (ES)] contributing to transport through the QD. Similar to the single-level model above we further assume an asymmetric broadening of the Fermi distribution in the leads with effective temperatures $T_S$~$<$~$T_D$ and a finite applied bias voltage $V_{QD}$~$>$~$0$. A schematic drawing of this model is shown in Fig.~\ref{fig04}(b). To describe a Coulomb resonance with completely inverted current flow, we furthermore assign different tunneling rates to the three different QD states. Considering higher rates for the two ES than for the GS (indicated by the widths of the arrows in Fig.~\ref{fig04}(b)) the net current flowing through the QD can be entirely negative as illustrated in Fig.~\ref{fig04}(c).
Indeed, a comparison of the experimental data in Fig.~\ref{fig02}(d) to the model in Fig.~\ref{fig04}(c) shows an apparently qualitative agreement.\\
Similar as for the single level model, values of $T_S$ and $T_D$ for the leads of the CNT QD can be determined. Fig.~\ref{fig04}(d) shows the extracted parameters as function of $V_{CD}$. $T_S$ and $T_D$ both increase approx. linearly with $V_{CD}$ and show a qualitatively similar behavior compared to the values extracted for the graphene QD analyzed by the single-level model (cf. Fig.~\ref{fig03}(d)).
The increase of $T_S$ and $T_D$ with $V_{CD}$ are measured to be $m_S$~$\approx$~49~K/V and $m_D$~$\approx$~30~K/V, respectively. Thus they differ only by a factor of 1.6 in contrast to 6.1 for the graphene QD (cf. Fig.~\ref{fig03}(d)).\\
If the tunnel rates of the three levels are assumed to be dependent on energy $E$, the three level model is furthermore capable to describe the sequence of Coulomb peaks shown in Fig.~\ref{fig04}(a). A corresponding sequence of calculated peaks for different broadening asymmetries ($T_S$~=~$T_D$ (black), $T_S$~=~2$T_D$ (blue) and $T_S$~=~4$T_D$) are shown in Fig.~\ref{fig04}(e). To obtain the illustrated traces, energy dependent tunneling rates fluctuating by over four orders of magnitude had to be assumed. The calculated traces are in qualitative agreement with the experimental data [(compare Figs.~4(a) and 4(e)].\\

The single-level and three-level transport model discussed above both rely on an asymmetric broadening of the source and drain leads as their most important assumption. The experimental evidence and the comparison to the model presented in Figs.~\ref{fig03} and \ref{fig04} show that this  broadening has to be induced by the interaction of the GNR based charge sensors with the QDs.
More insights can be gained from the quadratic dependency of the Coulomb peak current on $V_{CD}$ (see Fig.~\ref{fig03}(e)) which suggests a proportionality to the power dissipated in the CD. For typical $V_{CD}$ used in our experiments the dissipated power (which is most likely predominantly transferred to the CDs phonon bath) is on the order of hundreds of pW to tens of nW.
However, the transfer of the thermal energy from the graphene CDs to the QDs is hard to explain. While detector generated phonons are regarded as a relevant origin for back action in GaAs quantum dot devices~\cite{Granger2012}, phonon mediated energy transfer via the substrate is unlikely in carbon based devices due to the mismatch of the phonon spectra of graphene, CNTs and SiO$_2$.
It is more likely that energy is transferred by photons emitted by the current passing
through the etched graphene nanoribbon (GNR) based detectors~\cite{Molitor2010}.
Photon absorption in the leads of the QD devices could in turn lead to an effective broadening of its Fermi distribution.
Moreover, it is possible that the occupation of the stochastic QDs formed in the GNR~\cite{Molitor2010,Han2007,Todd2009} fluctuates. This would alter the potential landscape of the capacitively coupled QD and allow charge pumping effects. Averaged over time this would have the same effect as an effective increase in temperature. Finally, the source-drain broadening asymmetry assumed in the model is likely to be a direct consequence of the asymmetry in the device geometry (see Figs.~\ref{fig01}(a,b)) where one of the QD leads is closer to the CD than the other.

\section{Conclusion}
Back action effects of graphene nanoribbon based charge detectors (CDs) on graphene and carbon nanotube quantum dots (QDs) have been studied. We show that the current through the QDs is strongly influenced by the voltage applied to the CD. In both types of devices this voltage can lead to an enhanced transport through the QD but also to a reversal of the current. We employ a simple model considering transport only involving a single ground state which can explain the bias dependent behavior of isolated Coulomb peaks. A more advanced model taking into account excited state transport and state dependent tunneling rates proves suitable to furthermore describe longer sequences of Coulomb peaks. Importantly, both models assume an asymmetric broadening of the Fermi distribution in the leads of the QDs induced by the CD. Further experiments have to be conducted to obtain a better understanding of the mechanism of energy transfer between the charge detector and the QD. The reported study gives important information on the influence of graphene charge detectors on carbon based quantum dots which have to be considered when designing quantum dot circuits.

\section*{Acknowledgement}
We thank Federica Haupt, Johannes B\"ulte, Martin Leijnse, Wolfgang Belzig and Maarten Wegewijs for helpful discussions and preliminary calculations.
Support by the HNF, JARA Seed Fund and the DFG (SPP-1459 and FOR-912) are gratefully acknowledged.


\begin{thebibliography}{99}


\providecommand{\natexlab}[1]{#1}
\providecommand{\url}[1]{\texttt{#1}}
\expandafter\ifx\csname urlstyle\endcsname\relax
  \providecommand{\doi}[1]{doi: #1}\else
  \providecommand{\doi}{doi: \begingroup \urlstyle{rm}\Url}\fi



\bibitem{Petta2005}
J. R. Petta, A. C. Johnson, J. M. Taylor, E. A. Laird, A. Yacoby, M. D. Lukin, C. M. Marcus, M. P. Hanson, and A. C. Gossard.
\emph{Coherent Manipulation of Coupled Electron Spins in Semiconductor Quantum Dots.}
Science \textbf{309}, 2180 (2005).

\bibitem{Hayashi2003}
T. Hayashi, T. Fujisawa, H. D. Cheong, Y. H. Jeong, and Y. Hirayama.
\emph{Coherent Manipulation of Electronic States in a Double Quantum Dot.}
Phys. Rev. Lett. \textbf{91}, 226804

\bibitem{Gustavsson2006}
S. Gustavsson, R. Leturcq, B. Simovic, R. Schleser, T. Ihn, P. Studerus, and K. Ensslin
\emph{Counting Statistics of Single Electron Transport in a Quantum Dot. }
Phys. Rev. Lett. \textbf{96}, 076605 (2006).

\bibitem{Fujisawa2006}
T. Fujisawa, T. Hayashi, R. Tomita, and Y. Hirayama
\emph{Bidirectional Counting of Single Electrons.}
Science \textbf{312}, 1634 (2006).

\bibitem{Bluhm2010}
H. Bluhm, S. Foletti, I. Neder, M. Rudner, D. Mahalu, V. Umansky, and A. Yacoby.
\emph{Dephasing time of GaAs electron-spin qubits coupled to a nuclear bath exceeding 200 us. }
Nat. Phys. \textbf{7} 109 (2010).

\bibitem{Nowack2011}
K. C. Nowack, M. Shafiei, M. Laforest, G. Prawiroatmodjo, L. R. Schreiber, C. Reichl, W. Wegscheider, and L. M. K. Vandersypen.
\emph{Single-Shot Correlations and Two-Qubit Gate of Solid-State Spin.}
Science \textbf{333} 1269 (2011).

\bibitem{Shulman2012}
M. D. Shulman, O. E. Dial, S. P. Harvey, H. Bluhm, V. Umansky, and A. Yacoby.
\emph{Demonstration of Entanglement of Electrostatically Coupled Singlet-Triplet Qubits.}
Science \textbf{336} 202 (2012).

\bibitem{Elzermann2003}
J. M. Elzermann, R. Hanson, J. S. Greidanus, L. H. W. van Beveren, S. De Franceschi, L. M. K. Vandersypen, S. Tarucha, and L. P. Kouwenhoven.
\emph{Few-Electron quantum dot circuit with integrated charge read out. }
Phys. Rev. B \textbf{67} 161308 (2003).

\bibitem{Goetz2008}
G. G\"otz, G. Steele, W-J Vos, and L. P. Kouwenhoven.
\emph{Real Time Electron Tunneling and Pulse Spectroscopy in Carbon Nanotube Quantum Dots.}
Nano Lett. \textbf{8} 4039 (2008).

\bibitem{Shorubalko2008}
T. Shorubalko, R. Leturcq, A. Pfund, D. Tyndall, R. Krischek, S. Sch\"on, and K. Ensslin.
\emph{Self-Aligned Charge Read-Out for InAs Nanowire Quantum Dots.}
Nano Lett. \textbf{8} 382 (2008).

\bibitem{Choi2012}
T. Choi, T. Ihn, S. Sch\"on, and K. Ensslin.
\emph{Counting statistics in an InAs nanowire quantum dot with a vertically coupled charge detector.}
Appl. Phys. Lett. \textbf{100} 072110 (2012).

\bibitem{Guettinger2008}
J.~G\"uttinger, C.~Stampfer, S.~Hellm\"uller, F.~Molitor, T.~Ihn, and
  K.~Ensslin.
\newblock \emph{Charge detection in graphene quantum dots}.
\newblock Applied Physics Letters 93\penalty0 (21):\penalty0 212102 (2008).

\bibitem{Volk2013}
C.~Volk, C.~Neumann, S.~Kazarski, S.~Fringes, S.~Engels, F.~Haupt, A.~M\"uller,
  and C.~Stampfer.
\newblock \emph{Probing relaxation times in graphene quantum dots}.
\newblock Nature Communications \textbf{4}, \penalty0 1753 (2013).

\bibitem{Huertas-Hernando2006}
D. Huertas-Hernando, F. Guinea, and A. Brataas.
\newblock \emph{Spin-orbit coupling in curved graphene, fullerenes, nanotubes, and nanotube caps}.
\newblock Phys. Rev. B~\textbf{74}, 155426 (2006).

\bibitem{Min2006}
H. Min, J. E. Hill, N. A. Sinitsyn, B. R. Sahu, L. Kleinman, and A. H. MacDonald.
\newblock \emph{Intrinsic and rashba spin-orbit interactions in graphene sheets}.
\newblock Phys. Rev. B~\textbf{74}, 165310 (2006).

\bibitem{Trauzettel2007}
B. Trauzettel, D. V. Bulaev, D. Loss, and G. Burkard.
\newblock \emph{Spin qubits in graphene quantum dots}.
\newblock Nat. Phys.~\textbf{3}, 192 (2007).

\bibitem{Engels2013}
S.~Engels, P.~Weber, B.~Terr\'es, J.~Dauber, C.~Meyer, C.~Volk, S.~Trellenkamp, U.~Wichmann, and C.~Stampfer.
\newblock \emph{Fabrication of coupled graphenenanotube quantum devices}.
\newblock Nanotechnology \textbf{24}, \penalty0 035204 (2013).

\bibitem{Sols2007}
F. Sols, F. Guinea, and A. Castro Neto.
\emph{Coulomb Blockade in Graphene Nanoribbons.}
Phys. Rev. Lett. \textbf{99}, 166803 (2007).

\bibitem{Stampfer2011}
C. Stampfer, S. Fringes, J. G\"uttinger, F. Molitor, C. Volk, B. Terres, J. Dauber, S. Engels, S. Schnez, A. Jacobsen, S. Dr\"oscher, T. Ihn, and K. Ensslin.
\emph{Transport in Graphene Nanostructures.}
Frontiers of Physics \textbf{6}, 271 (2011).

\bibitem{Fringes2011}
S.~Fringes, C.~Volk, C.~Norda, B.~Terr\'{e}s, J.~Dauber, S.~Engels,
  S.~Trellenkamp, and C.~Stampfer.
\newblock \emph{Charge detection in a bilayer graphene quantum dot}.
\newblock Physica Status Solidi B \textbf{248}, \penalty0 2684 (2011).

\bibitem{Wang2010}
L.-J. Wang, G.~Cao, T.~Tu, H.-O. Li, C.~Zhou, X.-J. Hao, Z.~Su, G.-C. Guo,
  H.-W. Jiang, and G.-P. Guo.
\newblock \emph{A graphene quantum dot with a single electron transistor as an
  integrated charge sensor}.
\newblock Applied Physics Letters 97\penalty0 (26):\penalty0 262113 (2010).

\bibitem{Mueller2012}
T.~M\"uller, J.~G\"uttinger, D.~Bischoff, S.~Hellm\"uller, K.~Ensslin, and
  T.~Ihn.
\newblock \emph{Fast detection of single-charge tunneling to a graphene quantum
  dot in a multi-level regime}.
\newblock Appl. Phys. Lett. \textbf{101}, \penalty0 012104 (2012).

\bibitem{Guettinger2011a}
J.~G\"uttinger, J.~Seif, C.~Stampfer, A.~Capelli, K.~Ensslin, and T.~Ihn.
\newblock \emph{Time-resolved charge detection in graphene quantum dots}.
\newblock Phys. Rev. B \textbf{83}, \penalty0 165445 (2011).

\bibitem{Biercuk2006}
M. J. Biercuk, D. J. Reilly, T. M. Buehler, V. C. Chan, J. M. Chow, R. G. Clark, and C. M. Marcus.
\emph{Charge sensing in carbon-nanotube quantum dots on microsecond timescales.}
Phys. Rev. B \textbf{73}, 201402 (2006).

\bibitem{Zhou2012}
X. Zhou and K. Ishibashi.
\emph{Single charge detection in capacitively coupled integrated single electron transistors based on single-walled carbon nanotubes.}
Appl. Phys. Lett. \textbf{101}, 123506 (2012).

\bibitem{Churchill2009}
H. O. H. Churchill, A. J. Bestwick, J. W. Harlow, F. Kuemmeth, D. Marcos, C. H. Stwertka, S. K. Watson, and C. M. Marcus.
\emph{Electron–nuclear interaction in 13C nanotube double quantum dots.}
Nature Phys. \textbf{5} 321 (2009).

\bibitem{Durrer2008}
L. Durrer, T. Helbling, C. Zenger, A. Jungen, C. Stampfer, and C. Hierold.
\emph{SWNT growth by CVD on Ferritin-based iron catalyst nanoparticles towards CNT sensors.}
Sensors Actuators B \textbf{132}, 485 (2008).

\bibitem{Neumann2013}
C.~Neumann, C.~Volk, S.~Engels, and C.~Stampfer.
\newblock \emph{Graphene-based charge sensors}.
\newblock Nanotechnology \textbf{24}, \penalty0 444001 (2013).

\bibitem{Granger2012}
Granger, G. and Taubert, D. and Young, C.E. and Gaudreau, L. and Kam, A. and Studenikin, S.A. and Zawadzki, P. and Harbusch, D. and Schuh, D. and Wegscheider, W. and Wasilewski1, Z.R. and Clerk, A.A. and Ludwig, S. and Sachrajda, A.S.
\newblock \emph{Quantum interference and phonon-mediated back-action in lateral quantum-dot circuits}.
\newblock Nature Physics \textbf{8}, \penalty0 522 (2012).

\bibitem{Molitor2010}
F. Molitor, C. Stampfer, J. G\"uttinger, A. Jacobsen, T. Ihn, and K. Ensslin.
\emph{Energy and transport gaps in etched graphene nanoribbons.}
Semiconductor Science and Technology \textbf{25}, 034002 (2010).

\bibitem{Han2007}
M. Y. Han, B. \"Ozyilmaz, Y. Zhang, and P. Kim.
\emph{Energy Band Gap Engineering of Graphene Nanoribbons.}
Physical Review Letters \textbf{98}, 206805 (2007).

\bibitem{Todd2009}
K. Todd, H.-T. Chou, S. Amasha, and D. Goldhaber-Gordon.
\emph{Quantum Dot Behavior in Graphene Nanoconstrictions.}
Nano Letters \textbf{9}, 416--421 (2009).

\end{thebibliography}
\end{document}